\documentclass[a4paper,11pt]{article}

\usepackage{jheppub}
\usepackage{mathrsfs}
\usepackage{xspace}
\usepackage{times}
\usepackage{etoolbox}

\newcommand{\MS}{$\overline{\rm MS}$\xspace}
\newcommand{\OS}{OS\xspace}
\newcommand{\CP}{\textit{CP}\xspace}
\newcommand{\Tr}{{\rm Tr}}
\newcommand{\diag}{{\rm diag}}
\renewcommand{\Re}{\mbox{Re}}
\renewcommand{\Im}{\mbox{Im}}
\renewcommand{\vec}[1]{\mathbf{#1}}

\makeatletter
\patchcmd{\maketitle}{\@fpheader}{}{}{}
\makeatother

\preprint{MPP-2013-254}

\title{Basis invariant measure of \CP-violation\newline and renormalization}

\author{A. Hohenegger$^{a}$ and}
\emailAdd{andreas.hohenegger@uis.no}

\author{A. Kartavtsev$^{b}$}
\emailAdd{alexander.kartavtsev@mpp.mpg.de}

\affiliation[a]{University of Stavanger, Kjell Arholms gate 41, 4036 Stavanger, Norway}
\affiliation[b]{Max-Planck-Institut f\"ur Physik, F\"ohringer Ring 6, 80805 M\"unchen, Germany}

\arxivnumber{1309.1385}

\abstract{We analyze, in the context of a simple toy model, for which renormalization 
schemes the \CP-properties of bare Lagrangian and its finite part coincide. We show that 
this is the case for the minimal subtraction and on-shell schemes. The \CP-properties of 
the theory can then be characterized by \CP-odd basis invariants expressed in terms of
renormalized masses and couplings. For the minimal subtraction scheme we 
furthermore show that in \CP-conserving theories the \CP-odd basis invariants are zero 
at any scale but are not renormalization group invariant in \CP-violating ones.}

\notoc

\begin{document}

\maketitle
\flushbottom
\newpage

\section{\label{Introduction}Introduction}

Neutrino oscillations, i.e.~the experimental evidence for leptonic flavor-mixing, have established the existence of small but nonzero neutrino masses. Through a realization of the seesaw mechanism these can find a satisfying theoretical explanation which entails further interesting phenomenological consequences. In particular \CP-violating phases in the leptonic mixing open the possibility to explain the baryon asymmetry of the universe through the leptogenesis scenario \cite{Fukugita:1986hr}. Analogous to the complex phase in the Cabibbo-Kobayashi-Maskawa matrix, \CP-violating phases in the leptonic mixing may result from phases in vacuum expectation values of the Higgs fields or from complex Yukawa couplings. These phases will in general cause leptonic \CP-violation. However, not all of the phases are necessarily physical as they may be rotated away by weak basis transformations. Such rotations of the weak basis are in fact part of general \CP-transformations defined by the gauge sector of the theory. 
Therefore it is useful to discuss \CP-violating phenomena in terms of basis invariant quantities.

The strength of \CP-violation in a given model can be parametrized in terms of a 
few \CP-odd flavor-basis invariants which vanish if \CP is conserved. Originally 
they have been introduced in \cite{Jarlskog:1985ht} to provide a 
convention-independent measure of \CP-violation in the quark sector of the 
Standard Model. In \cite{Branco:2001pq,Branco:2006ce,Branco:2011zb} similar 
invariants have been constructed to parametrize \CP-violation in the leptonic 
sector of the Standard Model supplemented by heavy Majorana neutrinos 
\cite{Minkowski:1977sc,Yanagida:1979as,Mohapatra:1979ia,GellMann:1980vs}.
In a perturbative calculation \CP-violation manifests itself at loop level. The 
loop contributions are in general divergent and must be renormalized. Thus, we 
have to distinguish between bare and renormalized quantities. After 
renormalization the original Lagrangian can be represented as a sum of a basic 
Lagrangian, which has the same form as the bare one but contains only the 
renormalized quantities, and counterterms. Analyzing the basic Lagrangian
one can define the flavor-basis invariants characterizing its \CP-properties.
However, it is important to keep in mind that the \CP-properties of the basic 
Lagrangian may differ from those of the bare one. For instance, even if the 
the basic Lagrangian is \CP-conserving the counterterms may contain \CP-violation, 
such that the full theory is \CP-violating. 

In section \ref{BareAndBasic} we analyze for which renormalization schemes 
\CP-properties of the bare and basic Lagrangians coincide. For such schemes the 
strength of \CP-violation of the full theory can be characterized by the \CP-odd 
flavor-basis invariants expressed in terms of the renormalized masses and 
couplings. In section \ref{Running} we study properties of the these invariants 
under renormalization group running. We find that in \CP-conserving theories 
it is zero at any scale but is not renormalization group invariant in 
\CP-violating ones. Finally, in section \ref{Summary} we summarize our results.

\section{\label{BareAndBasic}CP-properties of the bare and basic Lagrangian}

To reduce the technical complications to a minimum here we use a simple toy model 
that has been used in \cite{Garny:2009rv,Garny:2009qn,Garny:2010nz,Garny:2010nj,
Hohenegger:2014cpa,Hohenegger:2014cpa} to study qualitative features of leptogenesis in the 
framework of non-equilibrium quantum field theory. The action is given by 
$S=\int d^4x {\cal L}$ and the Lagrangian of the model contains one complex and 
two real scalar fields:
\begin{align}
 \label{lagrangian}
 {\cal L} & = \frac12 \partial^\mu\psi_{0,i}\partial_\mu\psi_{0,i}
    - \frac12 \psi_{0,i} M^2_{0,ij} \psi_{0,j}
    + \partial^\mu \bar{b}\partial_\mu b 
    - \frac{h_{0,i}}{2!}\psi_{0,i} bb
    - \frac{h^*_{0,i}}{2!}\psi_{0,i} \bar{b}\bar{b}\,,
\end{align}
where $i,j=1,2$, the bar denotes complex conjugation and the subscript `0' denotes the bare fields, 
couplings and mass parameters. The real and symmetric mass matrix $\hat M_0^2$ mixes the 
two generations of real scalar fields $\psi_{0,i}$. The couplings $h_0$ take arbitrary 
complex values and can induce \CP-violation. Rephasing the complex field, we can always 
make one of the couplings real. On the other hand, the relative phase of the couplings is 
rephasing invariant. The renormalized fields, masses and couplings are related to the 
bare ones by
\begin{subequations}
\label{Renormalization}
\begin{align}
\label{FieldRenormalization}
\psi_{0,i}&=Z^\frac12_{\psi,ij}\psi_j\approx \psi_i+\frac12\delta Z_{\psi,ij}\psi_j\,,\\
\label{MassRenormalization}
M^2_{0,ij}&=M^2_{ij}+\delta M^2_{ij}\,,\\
\label{CouplingRenormalization}
h_{0,i}&=Z_{h,ij} h_{j}\approx h_{i}+\delta Z_{h,ij} h_{j}\,.
\end{align}
\end{subequations}
The matrix $Z_\psi$ is a general real matrix which is relevant for the renormalization 
of mixing fields \cite{Kniehl:1996bd,Pilaftsis:1997dr,Bouzas:2000np,Bouzas:2003ju}, 
and the matrix $Z_h$ is a general complex matrix. Rewritten in terms of the renormalized 
fields, masses and couplings the Lagrangian takes the form
\begin{align}
 \label{RenLagrangian}
 {\cal L} & = \frac12 \partial^\mu\psi_{i}\partial_\mu\psi_{i}
    - \frac12 \psi_{i} M^2_{ij} \psi_{j}
    + \partial^\mu \bar{b}\partial_\mu b 
    - \frac{h_{i}}{2!}\psi_{i} bb
    - \frac{h^*_{i}}{2!}\psi_{i} \bar{b}\bar{b} + \delta {\cal L}\,.
\end{align}
The counterterms read
\begin{align}
\label{Counterterms}
\delta {\cal L}&=\frac12 \partial^\mu \psi_i \Delta Z_{ij} \partial_\mu \psi_j
-\frac12 \psi_i \Delta M^2_{ij} \psi_j - \frac{\Delta h_{i}}{2!}\psi_{i} bb
- \frac{\Delta h^*_{i}}{2!}\psi_{i} \bar{b}\bar{b} \,,
\end{align} 
where we have introduced 
\begin{subequations}
\label{CountertermRelations}
\begin{align} 
\Delta Z_{ij} & \equiv \frac12 \delta Z_{\psi,ij}+ \frac12\delta Z^T_{\psi,ij}\,,\label{CountertermRelationsA}\\
\Delta M^2_{ij} & \equiv \delta M^2_{ij}+
\frac12 M^2_{ik} \delta Z_{\psi,kj}+\frac12 \delta Z^T_{\psi,ik} M^2_{kj} \,,\label{CountertermRelationsB} \\
\Delta h_{i} & \equiv \delta Z_{h,ij}h_j+\frac12 \delta Z^T_{\psi,ij}h_j\,.\label{CountertermRelationsC}
\end{align}
\end{subequations}
Let us begin with the analysis of the basic Lagrangian. Generically 
\CP-transformation turns a complex scalar field into its complex 
conjugate evaluated at $x=(x_0,-\vec{x})$ times an arbitrary phase: 
\begin{subequations}
\begin{align}
(CP) b(x_0,\vec{x}) (CP)^{-1}&=\beta \bar{b}(x_0, -\vec{x})\,,\\
(CP) \bar{b}(x_0,\vec{x}) (CP)^{-1}&=\beta^* b(x_0, -\vec{x})\,.
\end{align}
\end{subequations}
The complete \CP-transformation for the mixing scalar fields $\psi$ is found by splitting 
the Lagrangian into kinetic part and rest. The kinetic part is taken to define \CP and the 
complete \CP-tra\-n\-s\-formation therefore includes an internal (orthogonal) symmetry 
transformation $U_{ij}$ which leaves this term invariant,\footnote{Similarly, in the electroweak 
theory of the SM, the `generalized' \CP-transformation would be defined as a generalized 
symmetry transformation which leaves the kinetic- and gauge-part of the Lagrangian invariant 
\cite{Branco:1999cp}.}
\begin{align}
(CP) \psi_{i}(x_0,\vec{x}) (CP)^{-1}&=U_{ij}\psi_{j}(x_0, -\vec{x})\,.
\end{align}
The invariance properties of the remainder determine to which extent the Lagrangian violates \CP.
The internal symmetry transformation can be a flavor rotation or reflection,\footnote{Note that we use the term `flavor' throughout the paper to denote the generations of scalar fields of the toy-model.}
\begin{align}
\label{CFlavorTrafos}
U=\left(
\begin{tabular}{cc}
$c$ & $-s$\\
$s$ & $c$
\end{tabular} 
\right)\quad {\rm or}\quad 
U=\left(
\begin{tabular}{cc}
$c$ & $s$\\
$s$ & $-c$
\end{tabular} 
\right)\,,
\end{align}
where we have introduced $c\equiv \cos(\alpha)$ and $s\equiv \sin(\alpha)$ to 
shorten the notation. A product of a flavor rotation and reflection is again a 
reflection. Comparing the \CP-transformed action $S=(CP) S(CP)^{-1}$ 
with its original form 
we obtain the following conditions for \CP-in\-va\-ri\-an\-ce:
\begin{subequations}
\label{CSymmetryComditions}
\begin{align}
U^T_{im}M^2_{mn}U_{nj}&=M^2_{ij}\,,\\
\beta^2 U^T_{ik}\, h_{k}&=h^*_{i}\,.
\end{align}
\end{subequations}
If for a given set of couplings and mass parameters we can 
find $\beta$ and $U_{ij}$ such that conditions \eqref{CSymmetryComditions} 
are fulfilled then the Lagrangian is \CP-invariant.
In general, the mass matrix has nonzero off-diagonal elements. To simplify 
the analysis we perform a flavor rotation to the basis where it is 
diagonal, $M^2 = \diag(M_{1}^2,M_{2}^2)$. Assuming that $M_{1}^2 \neq M_{2}^2$, in 
 this basis, the first condition is fulfilled only for rotations by $\alpha=0,\pi$ and 
reflections about $\alpha/2=0,\pi/2$, i.e.~we have to consider only four choices of $U_{ij}$. 
The second of conditions \eqref{CSymmetryComditions} is equivalent to the requirement 
that the matrix $H_{ij}\equiv h_{i}h^*_{j}$ obeys $U^T_{im}H_{mn}U_{nj}=H^*_{ij}$. For 
$\alpha=0,\pi$ rotations this implies $H_{12}=H^*_{12}$. This equality holds if 
$\Im\,H_{12}=0$. For $\alpha=0,\pi$ reflections the second condition implies 
$H_{12}=-H^*_{12}$, which is fulfilled if $\Re\,H_{12}=0$. To analyze the special 
case of equal mass parameters, $M_{1}^2 = M_{2}^2$, we need the transformation rules 
for $\Im\,H_{12}$ and $\Re\,H_{12}$. Under a flavor rotation:
\begin{subequations}
\begin{align}
\Im\,H_{12} &\rightarrow \Im\,H_{12}\,,\\
\Re\,H_{12} &\rightarrow (c^2-s^2)\Re\,H_{12}
+cs (H_{22}-H_{11})\,.
\end{align}
\end{subequations}
Evidently, $\Im\,H_{12}$ is an invariant, while $\Re\,H_{12}$ can be made zero through 
a rotation by the angle
\begin{align}
\alpha=\frac12\arctan\frac{2 \Re\,H_{12}}{H_{11}-H_{22}} \,.
\end{align}
If the mass matrix is proportional to unity, then we can always rotate to the basis where 
$\Re\,H_{12}$ vanishes. Therefore, the Lagrangian is also \CP-invariant in this case. 
Summarizing the above, the basic Lagrangian \eqref{RenLagrangian} is \CP-invariant if 
either $\Im\,H_{12}=0$, $\Re\,H_{12}=0$ in the basis where the mass matrix is diagonal, 
or the mass matrix is proportional to unity. Let us now consider
\begin{align}
\label{JarlskogInv}
J\equiv\Im\,\Tr({H}{M}^3{H}^T{M})\,.
\end{align}
As can readily be verified, $J$ is invariant 
under the flavor transformations and, using \eqref{CSymmetryComditions} in a general basis, 
that it is \CP-odd. In the basis, in which the mass matrix is diagonal it takes the form 
\begin{align}
\label{Jarlskog}
J&=2\, \Im\,H_{12} \Re\,H_{12}M_{1}M_{2}(M^2_{2}-M^2_{1})\,.
\end{align}
Evidently, it vanishes if the theory is \CP-conserving. In other words, $J$ in \eqref{JarlskogInv},  
is a basis-in\-de\-pen\-dent measure of \CP-violation in the basic Lagrangian for the model 
under consideration. \CP-violating observables, such as \CP-violating parameters for the decays of $\psi_i$, are 
expected to be proportional to $J$ such that they vanish if $J=0$.

In order that the full renormalized Lagrangian be \CP-invariant, the sum of the renormalized 
masses and couplings and the corresponding counterterms must satisfy conditions similar to 
\eqref{CSymmetryComditions}:
\begin{subequations}
\label{MassAndCoupling}
\begin{align}
U^T_{im}(M^2_{mn}+\Delta M^2_{mn})U_{nj}&=(M^2_{ij}+\Delta M^2_{ij})\,,\\
\beta^2 U^T_{ik}\, (h_{k}+\Delta h_{k})&=(h_{i}+\Delta h_{i})^*\,.
\end{align}
\end{subequations}
The requirement of \CP-invariance of the kinetic term induces an additional condition,
\begin{align}
\label{WaveFunction}
U^T_{im}\Delta Z_{mn}U_{nj}&=\Delta Z_{ij}\,.
\end{align}
If \eqref{CSymmetryComditions} are fulfilled, then the resulting additional conditions for 
\CP-invariance of the full theory read:
\begin{subequations}
\label{RenCSymmetryComditions}
\begin{align}
\label{Cond1}
U^T_{im}\Delta Z_{mn}U_{nj}&=\Delta Z_{ij}\,,\\
\label{Cond2}
U^T_{im}\Delta M^2_{mn} U_{nj}&=\Delta M^2_{ij}\,,\\
\label{Cond3}
U^T_{im}\, \Delta H_{mn}\,U_{nj}&=\Delta H^*_{ij}\,,
\end{align}
\end{subequations}
where $\Delta H_{ij}\equiv h_i \Delta h^*_j+\Delta h_i h^*_j+\Delta h_i \Delta h^*_j$.
As before, we work in the flavor basis in which the mass matrix $M_{ij}^2$ is diagonal. 
The first and the second of 
the conditions \eqref{RenCSymmetryComditions} are trivially fulfilled for $\alpha=0,\pi$ 
rotations. Condition \eqref{Cond3} is then fulfilled if $\Im\,\Delta H_{12}=0$. 
For $\alpha=0,\pi$ reflections 
the first and second conditions are fulfilled only if both $\Delta Z$ and $\Delta M^2$ 
are also diagonal in the chosen basis. If this is the case the last condition then
demands $\Re\,\Delta H_{12}=0$. As explained above, if $M^2$ is proportional 
to unity then we rotate to the basis where $\Re\,H_{12}=0$. The full theory is 
\CP-conserving if $\Delta Z$ and $\Delta M^2$ are diagonal and $\Re\,\Delta H_{12}=0$ 
in this basis. 

Let us summarize for which $\Delta Z$, $\Delta M^2$ and $\Delta H$ the conditions
of \CP-invariance of the full theory reduce to those for the basic Lagrangian.
The first solution, $\Im\,H_{12}=0$, is sufficient for any choice of $\Delta Z$ and 
$\Delta M^2$, provided that $\Im\,\Delta H_{12}=0$. The second solution,
$\Re\,H_{12}=0$ in the basis where $M_{ij}^2$ is diagonal, exists only if 
$\Delta Z$ and $\Delta M^2$ are also diagonal in this basis, or become diagonal in 
this basis for $\Re\,H_{12}=0$, and if $\Re\,\Delta H_{12}=0$ 
in this basis. Finally, the third solution, $M^2 \propto 1$, is sufficient provided that 
$\Delta Z$ and $\Delta M^2$ are diagonal in the basis in which $\Re\,H_{12}=0$ and 
$\Re\,\Delta H_{12}=0$ in this basis. 
 
If the couplings and mass parameters in \eqref{JarlskogInv} are numerically equal for 
two different choices of $\Delta Z$, $\Delta M^2$ and $\Delta H$, i.e.~for two different 
renormalization schemes, then the values of $J$ are also equal. However, it is important 
to keep in mind that they correspond to two different bare Lagrangians and therefore we 
deal with two physically inequivalent theories. Consider for 
example the self-energy. The renormalized self-energy, $\Pi_{ij}$, is related to the 
unrenormalized one, $\Pi_{0,ij}$, by
\begin{align}
\label{RenSelfEnergy}
\Pi_{ij}(p^2)=\Pi_{0,ij}(p^2)-p^2\Delta Z_{ij}+\Delta M^2_{ij}\,.
\end{align}
In quantum field theory 
the self-energy contributes to physical observables. In particular, it shifts the pole masses 
and generates the self-energy \CP-violating parameters \cite{Garny:2009qn}. The divergent parts 
of the counterterms are fixed by the requirement that they cancel the divergent part of the 
self-energy. At the same time the finite part is restricted only by the requirement that the 
perturbative expansion must converge and differs in different renormalization schemes. Thus 
the explicit form of the self-energy is also different in different renormalization schemes. 
Therefore, if we would keep the couplings and mass parameters constant but change the 
renormalization scheme, the resulting values of the pole masses and \CP-violating parameters 
would also change.

We use dimensional regularisation. For the model considered here the one-loop unrenormalized self-energy is given 
by \cite{Garny:2009qn},
\begin{align}
\label{UnrenSelfEnergy}
\Pi_{0,ij}(p^2)=-\frac{\Re\,H_{ij}}{16\pi^2} B_0(p^2)\,,
\end{align}
where 
\begin{align}
B_0(p^2)=\Delta-\ln\frac{|p^2|}{\mu^2}+i\pi\theta(p^2)
\end{align}
is the usual two-point function \cite{tHooft1979365,Plumacher:1998ex} and 
$\Delta\equiv \epsilon^{-1}-\gamma+4\pi+2$ contains the divergent contribution.
We will also need the three-point functions. 
At one loop level they read:
\begin{subequations}
\label{GammaParticlesAndAntiparticles}
\begin{align}
\label{GammaParticles}
i\Gamma_{\psi_i bb}(p^2)=h_{0,i}^*+\frac{h_{0,i}}{16\pi^2}\sum_{j}h_{0,j}^{*2} C_0(p^2,0,M_j^2)\,,\\
\label{GammaAntiparticles}
i\Gamma_{\psi_i \bar{b}\bar{b}}(p^2)=h_{0,i}+\frac{h^*_{0,i}}{16\pi^2}\sum_{j}h_{0,j} C_0(p^2,0,M_j^2)\,,
\end{align}
\end{subequations}
where
\begin{align}
C_0(M_i^2,0,M_j^2)=\frac{1}{M_i^2}\biggl[
{\rm Li}_2\biggl(1+\frac{M_i^2}{M_j^2}\biggr)-\frac{\pi^2}{6}\biggr]\,,
\end{align}
is a complex-valued function and we have taken into account that 
$b$ is massless. Since $C_0$ is finite, the three-point functions
are finite as well. Note also that the three-point functions \eqref{GammaParticles} 
and \eqref{GammaAntiparticles} are different in the presence of \CP-violation.

Let us now consider the two most commonly used renormalization schemes, the 
\MS and \OS schemes. In both cases we define the counterterms in the basis 
where the matrix of the mass parameters is diagonal. In the \MS scheme
one introduces only those counterterms, which are required to cancel 
the divergencies:
\begin{subequations}
\label{MSbar}
\begin{align}
\Delta Z_{ij}&=0\,,\\
\Delta M^2_{ij}&=\frac{\Re\,H_{ij}}{16\pi^2}\Delta\,,\label{MSbarB} \\
\Delta H_{ij}&=0\,.
\end{align}
\end{subequations} 
Since $\Delta H_{ij}=0$ in this scheme, $\Im\,H_{12}=0$ is sufficient for \CP-invariance 
of the full theory. Furthermore, if $\Re\,H_{12}=0$ then the counterterm \eqref{MSbarB} 
is diagonal and the theory is also \CP-conserving in this case. For $M^2 \propto 1$ 
the form of the counterterms remains the same and the analysis is 
completely analogous. In the \OS scheme the renormalized self-energy is 
required to satisfy the following conditions:
\begin{subequations}
\label{renPrescriptionOS}
\begin{align}
	&\Pi_{ii}(p^2=M_i^2) = 0 \quad (i=1,2)\,,\\
	&\Pi_{ij}(p^2=M_i^2) = \Pi_{ij}(p^2=M_j^2) = 0 \,\,\, (i\not= j) \,,\\
	&\frac{d}{dp^2} \Pi_{ij}(p^2=M_i^2) = 0 \quad (i=1,2)\,.
\end{align}
\end{subequations}
Since the three-point functions \eqref{GammaParticles} and \eqref{GammaAntiparticles}
are in general different, it is impossible to choose $\Delta h_i$ such that 
$i\Gamma_{\psi_i bb}(M_i^2)=h_i^*$ and $i\Gamma_{\psi_i \bar{b}\bar{b}}(M_i^2)=h_i$
simultaneously. For this reason we choose it such that it renormalizes their \CP-symmetric
combination, 
\begin{align}
	i\Gamma^*_{\psi_i bb}(M_i^2)+ i\Gamma_{\psi_i \bar{b}\bar{b}}(M_i^2)=2 h_i\,.
\end{align}
The resulting counterterms read:
\begin{subequations}
\label{OS}
\begin{align}
\label{OSZ}
\Delta Z_{ij}&=\frac{\Re\,H_{ij}}{16\pi^2}\frac{\ln(M_i^2/M_j^2)}{M_i^2-M_j^2}\,,\\
\label{OSM}
\Delta M^2_{ij}&=\frac{\Re\,H_{ij}}{16\pi^2}\biggl[\Delta
-\frac{M_i^2\ln(M_j^2/\mu^2)-M_j^2\ln(M_i^2/\mu^2)}{M_i^2-M_j^2}\biggr]\,,\\
\Delta H_{ij}&=-\frac{1}{16\pi^2}\sum\limits_n\bigl[H_{in} H^*_{nj} \Re\, C_0(M_j^2,0,M_n^2)
+H^*_{in} H_{nj}\Re\, C_0(M_i^2,0,M_n^2)\bigr]\,.
\end{align}
\end{subequations} 
Since $\Im\,\Delta H_{12}=0$ for $\Im\,H_{12}=0$, this condition is sufficient for 
\CP-invariance of the full theory. If $\Re\,H_{12}=0$ then both $\Delta Z$ and 
$\Delta M^2=0$ are diagonal. Furthermore, in this case $\Re\,\Delta H_{12}=0$ and therefore the 
theory is \CP-conserving. For $M^2 \propto 1$ we obtain, taking the 
limit $M_j^2=M_i^2 = M^2$ in \eqref{OSZ} and \eqref{OSM},
\begin{subequations}
\label{OSEqMasses}
\begin{align}
\Delta Z_{ij}&=\frac{\Re\,H_{ij}}{16\pi^2}\frac{1}{M^2}\,,\\
\Delta M^2_{ij}&=\frac{\Re\,H_{ij}}{16\pi^2}\bigl(\Delta-\ln(M^2/\mu^2)+1\bigr)\,.
\end{align}
\end{subequations}
Since the flavor properties of \eqref{OSEqMasses} are determined by flavor 
properties of the overall factor $\Re\,H_{ij}$, we can always rotate to the basis where 
$\Re\,H_{12}=0$. In this basis both $\Delta Z$ and $\Delta M$ are diagonal and, as 
before, $\Re\,\Delta H_{12}=0$. Therefore, the theory is again \CP-conserving.
In other words, for the \MS and \OS renormalization 
schemes the definition \eqref{JarlskogInv} which characterizes \CP-properties of the basic 
Lagrangian can be used as a basis-invariant measure of \CP-violation in the full theory.

For illustrational purposes let us present a simple example where the full theory is 
\CP-violating even though basic Lagrangian is \CP-conserving. We choose
\begin{subequations}
\label{MSbarModified}
\begin{align}
\Delta Z_{ij}&=0\,,\\
\Delta M^2_{ij}&=\frac{\Re\,H_{ij}}{16\pi^2}\Delta+\Delta {\cal M}^2_{ij}\,,
\end{align}
\end{subequations} 
where, in the basis in which the mass matrix is diagonal, $\Re H_{12}=0$ and $\Delta {\cal M}^2_{ij}$ 
is a finite matrix with nonzero off-diagonal elements. For this choice $J=0$ but the 
con\-di\-ti\-on \eqref{Cond1} is violated and therefore the full theory is expected to be 
\CP-violating. To convince ourselves that this is indeed the case we can shift $\Delta {\cal M}^2_{ij}$ 
to the mass term of the basic Lagrangian. This transformation does not change the bare 
Lagrangian and therefore we deal with physically the same theory. After the transformation 
we have \MS counterterms and finite Lagrangian with a non-diagonal mass matrix. In the basis 
where the new mass matrix is diagonal $\Re H_{12}$ is no longer zero and therefore $J\neq 0$, as 
expected.

Above we have studied the conditions under which the full theory is \CP-invariant 
provided that the basic Lagrangian is \CP-invariant. However, one should keep in 
mind that there is also the possibility of exact cancellation such that the full 
theory is {\CP}-conserving even though both the basic Lagrangian and counterterms
are \CP-violating. For instance, for the choice of counterterms made in 
\eqref{MSbarModified} this would be the case if the matrix of the mass parameters 
in the basic Lagrangian has the form $M^2_{ij}=M^2\delta_{ij}- 
\Delta {\cal M}^2_{ij}$. In such a case perturbation theory at finite loop-order 
can result in {\CP}-violating quantities and also $J\neq 0$ even though the full 
theory is \CP-conserving.

\section{\label{Running}Renormalization group running}

Because the renormalization group running does not change the bare Lagrangian, 
the \CP-properties of the full theory are RG-invariant. On the other hand, it is 
not obvious that the running does not modify the \CP-properties of the counterterms 
and, consequently, also the \CP-properties of the basic Lagrangian.

In this section we derive renormalization group equations (RGE's) for the parameters 
of the theory and verify that they preserve the \CP-properties of the basic Lagrangian. In $D=4-2\epsilon$ dimensions
\begin{align}
 {\cal L} & = \frac12 \partial^\mu\psi_{0,i}\partial_\mu\psi_{0,i}
    - \frac12 \psi_{0,i} M^2_{0,ij} \psi_{0,j}
    + \partial^\mu \bar{b}\partial_\mu b 
    - \mu^\epsilon\frac{h_{0,i}}{2!}\psi_{0,i} bb
    - \mu^\epsilon\frac{h^*_{0,i}}{2!}\psi_{0,i} \bar{b}\bar{b}\,.
\end{align}
We work within the minimal subtraction scheme in which the counterterms are given by \eqref{MSbar} with $\Delta =\epsilon^{-1}$ (because the theory 
parameters in a given renormalization scheme can always be mapped to the parameters 
in the minimal subtraction scheme results of this section generalize to other schemes 
as well). 

The renormalisation group equations follow from the requirement that 
\begin{subequations}
\label{RGinitial}
\begin{align}
\mu\frac{d}{d\mu}(M^2_{0,ij})=
\mu\frac{d}{d\mu}(M^2_{ij}+\delta M^2_{ij})&=0\,,\\
\mu\frac{d}{d\mu}(\mu^\epsilon h_{0,i})=
\mu\frac{d}{d\mu}(\mu^\epsilon Z_{h,ij}h_j)&=0\,,
\end{align}
\end{subequations}
where $\delta M^2_{ij}$ and $Z_{h,ij}$ are the mass and coupling counterterms 
introduced above and which have to be determined by solving \eqref{MSbar}. Relations 
\eqref{CountertermRelations} are fulfilled in 
particular for $\delta Z_\psi=\delta Z_h=0$. This solution is not unique and others are 
possible which lead to different variants of the RGE's which are related by flavor rotations. Solving for 
$\mu\, d M^2/d\mu$ and $\mu\, d H/d\mu$ and taking the limit $\epsilon \rightarrow 0$ 
we obtain the RG-equations for masses and couplings:
\begin{subequations}
\label{RGEquation}
\begin{align}
\label{RGEquationMass}
\frac{d M^2_{ij}}{dt}&= \Re H_{ij}\,,\\
\label{RGEquationCoupling}
\frac{d H_{ij}}{dt}&=0\,,
\end{align}
\end{subequations}
where $t\equiv \ln(\mu^2/\mu^2_0)/(16\pi^2)$. They have the explicit solutions $H_{ij}(t) = H_{ij}(0)$ and 
\begin{align}
\label{MassScaleDependence}
M^2_{ij}(t)=M^2_{ij}(0)+\Re\,H_{ij}\cdot t\,.
\end{align}

Let us assume for a moment that at $t=0$ the basic Lagrangian is \CP-invariant.
As has been discussed above there are three possibilities. First, this is the case 
if $\Im H_{12}=0$. Since $H_{ij}$ is scale-independent $\Im H_{12}$ remains zero and 
therefore the basic Lagrangian remains \CP-invariant. The second possibility is 
$\Re H_{12}=0$. In this case the mass matrix remains diagonal at any scale. Since 
$H_{ij}$ is sale-independent the condition $\Re H_{12}=0$ is fulfilled for any 
$t$ and the basic Lagrangian remains \CP-invariant. Third, if $M^2_{ij}=
M^2\delta_{ij}$ at $t=0$ then we can rotate to the basis where $\Re H_{12}=0$
without changing the matrix of mass parameters. In the new basis $M^2_{ij}(t)$
is diagonal (though no longer proportional to unity for $t\neq 0$) 
and $\Re H_{12}=0$. Therefore, the basic Lagrangian remains \CP-conserving at any 
scale. This implies that renormalization group running does not change \CP-properties 
of the basic Lagrangian. For the basis-invariant measure of \CP-violation we find to
leading order in the couplings 
\begin{align}
\label{Jmudep}
J(t)\approx J(0)\left[1+
\frac{M^2_{2}(0)H_{11}+M^2_1(0)H_{22}}{2M^2_1(0)M^2_2(0)}
\cdot t\,\right]\,,
\end{align}
where we have assumed that the mass matrix $M^2_{ij}$ is diagonal at $t=0$.
This expression reflects that if $J=0$ at $t=0$ then it remains zero at any 
scale. On the other hand, from \eqref{Jmudep} it follows that the \CP-odd basis 
invariants are not renormalization group invariant in \CP-violating theories.

To conclude this section let us  note that the mass matrix 
\eqref{MassScaleDependence} can be diagonalized by a finite flavor transformation, 
$M^2\rightarrow U^T M^2 U$, which also transforms the couplings, $H\rightarrow U^T H U$.
This is referred to as `run and diagonalize' approach. On the other hand, one could 
pursue the `diagonalize and run' approach by requiring that as $t\rightarrow t+dt$ the 
mass matrix is brought to the diagonal form by an infinitesimally small flavor transformation, 
such that it remains diagonal at any scale. Combined with \eqref{RGEquationMass} this requirement 
gives $d\alpha/dt=\Re H_{12}/(M_2^2-M_1^2)$ for the derivative of the rotation angle, where 
$M$ and $H$ now denote the masses and couplings in the new basis. This gives for the derivatives
of the latter
\begin{subequations}
\label{RGEquation1}
\begin{align}
\label{RGEquationMass1}
\frac{d M^2_{ij}}{dt}&= \delta^{ij}\cdot \Re H_{ij}\,,\\
\label{RGEquationCoupling1}
\frac{d H_{ij}}{dt}&=\frac{\Re H_{12}}{M_2^2 - M_1^2}
\left(\begin{array}{cc}
-2 \Re H_{12} & H_{11} - H_{22} \\
 H_{11} - H_{22} & 2 \Re H_{12}
\end{array}\right)\,.
\end{align}
\end{subequations}
An alternative derivation of \eqref{RGEquation1} is presented in Appendix \ref{DiagonalizeRun}.
Note that because \eqref{RGEquation} and \eqref{RGEquation1} are equivalent by construction
they give (in the basis where the mass matrix is diagonal) the same results for the masses 
and couplings and therefore the same result for the scale-dependence of the \CP-odd basis-invariant, 
see \eqref{Jmudep}.

\section{\label{Summary}Summary}

To summarize, we have analyzed for which renormalization schemes \CP-properties 
of the bare and basic Lagrangians coincide. Since for the same couplings and mass 
parameters of the basic Lagrangian, which determine the value of the \CP-odd 
flavor invariant $J$, we can choose different renormalization schemes and 
therefore different counterterms (which would imply that the corresponding bare 
theories differ), the latter can induce \CP-violation even if $J=0$. However, 
for the two most commonly used schemes, the \MS and \OS schemes, the condition 
$J=0$ is sufficient to ensure that the full theory is \CP-conserving.

Because renormalization group running leaves the bare Lagrangian invariant it 
also does not change its \CP-properties. Therefore if the theory is \CP-conserving 
at the initial scale it remains \CP-conserving at other scales. Furthermore, we 
have found that (at least for the considered here toy model) renormalization group running 
also does not change \CP-properties of the basic Lagrangian and of the counterterms. 
Thus if $J$ is zero at the initial scale it remains zero at other scales. On the 
other hands if the theory is \CP-violating then $J$ depends on the scale. In other 
words, it is flavor-basis invariant but not RG invariant.

\begin{appendix}

\section{\label{DiagonalizeRun} Diagonalize and run approach}

In this appendix we pursue an alternative derivation of the renormalization group 
equations that is based on a parametrization of the renormalisation prescription which differs 
slightly from that of equation \eqref{Renormalization}. This prescription simplifies the 
computation of RGE's which automatically keep the mass matrix diagonal. We then analyze the 
\CP-properties in terms of the \CP-odd basis-invariant evaluated in the mass-diagonal basis.

To this end we use the minimal general parametrization
of the counterterms  
\cite{Bouzas:2000np,Bouzas:2003ju}:
\begin{subequations}
\label{RenormalizationGeneralized}
\begin{align}
\label{FieldRenormalizationGeneralized}
\psi_{0,i}&=(U Z^\frac12)_{ij}\psi_j\,,\\
\label{MassRenormalizationGeneralized}
M^2_{0,ij}&=U_{m,ik}^T (M^2_{kl}+\delta M^2_{kl}) U_{m,lj}\,,\\
\label{CouplingRenormalizationGeneralized}
h_{0,i}&=\mu^{\epsilon} Z_{h,ij} h_{j}\,,
\end{align}
\end{subequations}
where by means of polar decomposition we represent 
$Z^{\frac12}_\psi=U Z^{\frac12}$ with $U$ and $Z^{\frac12}$ being 
real orthogonal and symmetric matrices respectively.
In \eqref{RenormalizationGeneralized} we also require that the matrix $U_m$ is real and orthogonal and that the mass-matrix and $\delta M^2$ satisfies $[M^2,\delta M^2]=0$. Thereby \eqref{MassRenormalizationGeneralized} represents a minimal parametrization of a general transformation of a diagonalizable mass matrix \cite{Bouzas:2000np}.
With $U\approx   1-\delta U$, $U_m\approx  1-\delta U_m$, $Z=1+\delta Z$ and $Z_h=1+\delta Z_h$, we obtain
\begin{subequations}
\label{RenormalizationGeneralizedInfinitesimal}
\begin{align}
\label{FieldRenormalizationGeneralizedInfinitesimal}
\psi_{0,i}&\approx \psi_i+\frac12\delta Z_{ij}\psi_j -\delta U_{ij}\psi_j \,,\\
\label{MassRenormalizationGeneralizedInfinitesimal}
M^2_{0,ij}&\approx M^2_{ij}+\delta M^2_{ij} + [\delta U_m ,M^2]_{ij}\,,\\
\label{CouplingRenormalizationGeneralizedInfinitesimal}
h_{0,i}&\approx \mu^{\epsilon}( h_{i}+\delta Z_{h,ij} h_{j})\,,
\end{align}
\end{subequations}
where $\delta U$ and $\delta U_m$ are real anti-symmetric matrices. Instead of \eqref{CouplingRenormalizationGeneralizedInfinitesimal} we can also write $H_0\approx\mu^{2\epsilon}(H+\delta H)$, which defines $\delta H \equiv \delta Z_h H + H \delta Z_h^\dagger$.
These expressions are to be compared to \eqref{Renormalization}. 
In this parametrization equations \eqref{RGinitial} take the form 
\begin{subequations}
\label{RGinitial1}
\begin{align}
\label{RGinitial1Mass}
\mu\frac{d}{d\mu}(M^2_{ij}+\delta M^2_{ij}+[\delta U_m,M^2]_{ij})&=0\,,\\
\label{RGinitial1Coupling}
\mu\frac{d}{d\mu}(\mu^\epsilon Z_h^{ij}h_j)&=0\,.
\end{align}
\end{subequations}

Inserting relations \eqref{RenormalizationGeneralizedInfinitesimal} into the bare Lagrangian \eqref{lagrangian} and comparing with \eqref{RenLagrangian} and \eqref{Counterterms} reveals the relations to $\Delta M^2$, $\Delta h$ and $\Delta Z$, modifying \eqref{CountertermRelations}:
\begin{subequations}
\label{CountertermRelationsGeneralized}
\begin{align} 
\Delta Z_{ij} & = \delta Z_{ij}\,,
\label{CountertermRelationsAGeneralized}\\
\Delta M^2_{ij} & = \delta M^2_{ij}+
\frac12 M^2_{ik} \delta Z_{kj}+\frac12 \delta Z_{ik} M^2_{kj} +[\delta U + \delta U_m , M^2]_{ij} \,,\label{CountertermRelationsBGeneralized} \\
\Delta h_{i} & = \mu^\epsilon(\delta Z_{h,ij}+\frac12 \delta Z_{ij} + \delta U_{ij})h_j\,.\label{CountertermRelationsCGeneralized}
\end{align}
\end{subequations}
Deriving RGE's involves solving relations \eqref{CountertermRelationsGeneralized} for $\delta M^2$, $\delta Z_h$, $\delta Z$ (and $\delta U$, $\delta U_m$) such that we can express the bare parameters \eqref{RenormalizationGeneralizedInfinitesimal} in terms of renormalized ones and $\mu$.
To compute the renormalization group we choose MS-scheme counterterms as in the main text. As in \eqref{MSbar}, we use $\Delta Z = \delta Z =0$ and $\Delta h =0$, but $\Delta M^2 = \Re H/(16\pi^2\epsilon)$. 
Different RGE's are obtained by making different additional assumptions for $\delta U$ and $\delta U_m$. Choosing $\delta U = 0$, we obtain from \eqref{CountertermRelationsBGeneralized}:
\begin{align}
\delta M^2_{ij} + [\delta U_m ,M^2]_{ij}&=\Delta M^2_{ij}\,,
\end{align}
which has nonzero off-diagonal elements and
which we can insert in \eqref{RGinitial1Mass} without solving for $\delta U_m$ itself.
Similarly, from \eqref{CountertermRelationsCGeneralized} we get $\delta Z_h = 0$ 
and therefore $Z_h=1$ in \eqref{RGinitial1Coupling}.
This leads again to the result obtained in \eqref{RGEquation} and \eqref{MassScaleDependence}.
The mass-matrix acquires off-diagonals during RG-evolution unless $\Re H_{12} = 0$. The anomalous dimension of the fields, 
\begin{align}
\gamma_{ij} &\equiv \mu\, {d\delta Z_{\psi,ij}}/{d\mu},
\end{align}
is given by $\gamma_{ij} = 0$.

A second possibility to solve \eqref{CountertermRelationsGeneralized} consists in choosing 
$\delta U_m =0$. In this case the requirement $[M^2,\delta M^2]=0$ is fulfilled 
(assuming a diagonal basic mass matrix $M^2$) only if $\delta M^2$ is diagonal. From \eqref{RGinitial1Mass}
it then follows that  a diagonal mass matrix will always stay diagonal under RG-evolution with this choice.
As can be inferred from \eqref{CountertermRelationsBGeneralized}  the off-diagonals of 
$\Delta M^2$ have to be absorbed into $\delta U$ in this case.
Since $\delta U$ is anti-symmetric, in the basis where $M^2$ is diagonal $[\delta U,M^2]$ is symmetric with vanishing diagonals and
\begin{align}
\label{CommuatorDeltaUM2}
[\delta U,M^2]_{12} &= [\delta U,M^2]_{21} =\delta U_{12}(M_{2}^2-M_{1}^2)\,.
\end{align}
It follows with \eqref{CountertermRelationsBGeneralized} that 
\begin{subequations}
\begin{align}
\delta M_{ij}^2 =&\delta_{ij}\, \Delta M_{ii}^2\,,\\
\label{CountertermdeltaUij}
\delta U_{12} = & -\delta U_{21} = \frac{\Delta M_{12}^2}{M_2^2-M_1^2}\,,
\end{align}
\end{subequations}
where we used MS-scheme counterterms $\delta Z = \Delta Z =0$ again. Furthermore, with $\Delta h =0$, we get
$\delta Z_h=-\delta U$ and therefore
\begin{align}
\delta H & = -[\delta U,H] = -\delta U_{12} 
\left(\begin{array}{cc}
-2 \Re H_{12} & H_{11} - H_{22} \\
 H_{11} - H_{22} & 2 \Re H_{12}
\end{array}\right)
\,,
\end{align}where we used the definition of $\delta H$, \eqref{CountertermRelationsBGeneralized} and the fact that $\delta U$ is anti-symmetric. Using these relations in \eqref{RGinitial1}, solving these systematically neglecting higher orders in the couplings, and taking the limit $\epsilon\rightarrow 0$ we get:
\begin{subequations}
\label{RGEquationDiagonal}
\begin{align}
\label{RGEquationMassDiagonal}
\mu \frac{d M^2_{ij}}{d\mu}&=\delta_{ij}\frac{H_{ij}}{8\pi^2}\,,\\
\label{RGEquationCouplingDiagonal}
\mu \frac{d H_{ij}}{d\mu}&=\frac{\Re H_{12}}{8\pi^2 (M_2^2 - M_1^2)}
\left(\begin{array}{cc}
-2 \Re H_{12} & H_{11} - H_{22} \\
 H_{11} - H_{22} & 2 \Re H_{12}
\end{array}\right)\,.
\end{align}
\end{subequations}
The anomalous dimension of the fields may be obtained from \eqref{FieldRenormalizationGeneralizedInfinitesimal} and \eqref{CountertermdeltaUij}:
\begin{align}
\delta Z_{\psi,ij} & =\delta Z_{ij} -2\delta U_{ij}=-\frac{\Re H_{ij}}{8\pi^2 (M_j^2-M_i^2)\epsilon}\,,
i\neq j\,.
\end{align}
This results in 
\begin{align}
\gamma_{ij} & = \frac{\Re H_{12}}{4\pi^2 (M_2^2-M_1^2)}
\left(\begin{array}{cc}
0 & 1 \\
-1 & 0
\end{array}\right)
\,.
\end{align}
It describes how the fields corresponding to the eigenvalues of the mass-matrix change their identity as the scale changes since these behave under RG-running as $\mu d\psi_i/d\mu = -\frac12 \gamma_{ij}\psi_j$.

For $M_2^2=M_1^2$ it is apparent from \eqref{CommuatorDeltaUM2} and \eqref{CountertermRelationsBGeneralized} that there is in general no solution to the counter-term relations with diagonal $\delta M^2$. Therefore this case must be treated separately. We may rotate to the basis in which $\Re H_{12}=0$. 
In this basis the solutions for the counter-terms are then given by ($\delta Z = \Delta Z =0$):
\begin{align}
\delta M_{ij}^2 =& \delta_{ij}\,\Delta M_{ii}^2\,,\\
\label{CountertermdeltaUijEqualM}
\delta U_{12} = & -\delta U_{21} = 0\,,\\
\delta Z_h = & 0\,.
\end{align}
Using these relations in \eqref{RGinitial1}, solving for the derivatives of the renormalized quantities and finally taking the limit $\epsilon\rightarrow 0$ yields
\begin{subequations}
\label{RGEquationDiagonalEqualM}
\begin{align}
\label{RGEquationMassDiagonalEqualM}
\mu \frac{d M^2_{ij}}{d\mu}&=\delta_{ij}\frac{H_{ij}}{8\pi^2}\,,\\
\label{RGEquationCouplingDiagonalEqualM}
\mu \frac{d H_{ij}}{d\mu}&=0
\,.
\end{align}
\end{subequations}

In fact we did not have to choose $\delta U_m =0$ above. If we solve \eqref{CountertermRelationsBGeneralized} for $\delta M^2$ and use the requirement $[M^2,\delta M^2]=0$, we find that $\delta U' \equiv \delta U +\delta U_m$ is fixed in terms of basic quantities and the counterterms $\Delta M^2$ and $\Delta Z$. Equations \eqref{CountertermRelationsGeneralized} therefore become
\begin{subequations}
\begin{align} 
\Delta Z_{ij} & = \delta Z_{ij}\,,\nonumber\\
\Delta M^2_{ij} & = \delta M^2_{ij}+
\frac12 M^2_{ik} \delta Z_{kj}+\frac12 \delta Z_{ik} M^2_{kj} +[\delta U' , M^2]_{ij} \,,\nonumber \\
\Delta h_{i} & = (\delta Z_{h,ij}' + \frac12 \delta Z_{ij} + \delta U_{ij}')h_j\,,\nonumber
\end{align}
\end{subequations}
where we introduced $\delta Z_{h}'=\delta Z_{h}-\delta U_{m}$. We know from the previous considerations that, once we choose the counterterms $\Delta Z$, $\Delta M^2$ and $\Delta h$ using the MS-scheme renormalization conditions, $\delta Z$, $\delta M^2$, $\delta Z_h'$ and $\delta U'$ are completely fixed by these equations. The quantity $\delta U_m$ can however be varied freely as long as $\delta Z_{h}$ and $\delta U$ are varied simultaneously so as to compensate the change. The (anti-symmetric) changes in the matrices $\delta U_m$, $\delta Z_h$ and $\delta U$ affect the mass-matrix, couplings and bare fields respectively. One can show from the requirement that the bare quantities stay invariant that this anti-symmetric matrix which depends on a single 
parameter transforms the basic quantities as a rotation which can for instance be used to diagonalize $M^2$. In this representation the choice of basis used in arguments above appears as a degree of freedom in the renormalization prescription which leaves the counterterms unchanged. With respect to the RG-running derived above we can therefore equivalently use a prescription in which the mass matrix develops off-diagonals, such as that given in \eqref{Renormalization}, to fix the finite parts of the 
counterterms and diagonalize it afterwards (run-and-diagonalize approach).

Let us now study the evolution of the \CP-odd basis-invariant under RGE-evolution. Since the mass-matrix stays diagonal in this scheme we may use \eqref{Jarlskog}. Differentiating with respect to $\mu$  we find:
\begin{align}
\mu\frac{d}{d\mu}J = & 2\, \Im\,\beta_{12} \Re\,H_{12}M_{1}M_{2}(M^2_{2}-M^2_{1})\nonumber\\
&+2\, \Im\,H_{12} \Re\,\beta_{12}M_{1}M_{2}(M^2_{2}-M^2_{1})\nonumber\\
& -2\, \Im\,H_{12} \Re\,H_{12}M_{1}M_{2}(\gamma_{m,22}^2-\gamma_{m,11}^2)+\ldots\,,
\end{align}
where $\beta\equiv \mu d H/d\mu$, $\gamma_m\equiv -\mu d M^2/d\mu$ and the 
ellipses indicate terms proportional to the derivatives of $M_1$ and $M_2$, which have 
the same \CP-properties as $J$ itself.
We first consider the case $M_2^2\neq M_1^2$. Since $\beta\propto \Re H$, $\Im\beta_{12}=0$ and the first term vanishes identically. The second and third term vanish if either  $\Im H_{12}=0$ or $\Re\, \beta_{12}=0$, which, according to \eqref{RGEquationCouplingDiagonal}, is the case if it was the case for $\mu = \mu_0$. Therefore, $J$ remains zero if it has been zero initially. For $M_2^2 (\mu_0)=M_1^2 (\mu_0)$ we work in the basis in which $\Re H_{12} = 0$. According to \eqref{RGEquationCouplingDiagonalEqualM}, the first two terms vanish. Since $\Re H_{12}$ stays zero, the last term vanishes as well and $J=0$, even though the  eigenvalues of $M$ evolve under RG-running.

\end{appendix}

\bibliographystyle{JHEP}
\providecommand{\href}[2]{#2}\begingroup\raggedright\endgroup


\end{document}